\documentclass[aps,preprint]{revtex4}

\usepackage{tikz}
\usepackage{calc}
\usetikzlibrary{shapes,shapes.geometric,arrows,positioning,decorations,decorations.pathmorphing,chains,automata,calc,patterns}

\newcommand{\stargraph}[3][black]{\begin{tikzpicture}[node distance=0.5cm,baseline={-0.5*height("$=$")}]
  \node at (360:0mm) (c) {};
  \foreach \n in {1,...,#2}{
    \node at ({\n*360/#2}:#3cm) (n\n) {};
    \draw (c.center)--(n\n.center);
    \draw[fill=#1] (n\n) circle[radius=#3*4.16pt, color=black];
  \draw[fill=#1] (c) circle[radius=#3*4.16pt, color=black];
  }
\end{tikzpicture}}

\newcommand{\twonodes}[1][black]{\begin{tikzpicture}[node distance=0.45cm,baseline={0.2cm-0.5*height("$=$")}]
  \node (1) {};
  \node (2) [above of=1] {};
  \draw (1.center) to (2.center);
  \draw[fill=#1] (1) circle[radius=1.25pt, color=black];
  \draw[fill=#1] (2) circle[radius=1.25pt, color=black];
\end{tikzpicture}}

\newcommand{\drawpath}[2][black]{\begin{tikzpicture}[node distance=0.5cm,baseline={0.15cm-0.5*height("$=$")}]
  \node (1) {};
  \foreach \n in {2,...,#2}{
	\ifodd\n
      \path (\the\numexpr\n-1) ++(-60:0.5) node (\n) [draw,white,fill=white] {}; 
	\else
      \path (\the\numexpr\n-1) ++(60:0.5) node (\n) [draw,white,fill=white] {}; 
	\fi
	}
    \foreach \n in {2,...,#2}
      \draw (\the\numexpr\n-1.center) to (\n.center);
  \foreach \n in {1,...,#2}
    \draw[fill=#1] (\n) circle[radius=1.25pt, color=black];
\end{tikzpicture}}

\newcommand{\drawpathfoot}[1]{\begin{tikzpicture}[node distance=0.25cm,baseline={0.1cm-0.5*height("$=$")}]
  \node (1) {};
  \foreach \n in {2,...,#1}{
	\ifodd\n
	  \node (\n) [below right of=\the\numexpr\n-1] {};
	\else
	  \node (\n) [above right of=\the\numexpr\n-1] {};
	\fi
	}
    \foreach \n in {2,...,#1}
      \draw (\the\numexpr\n-1.center) to (\n.center);
  \foreach \n in {1,...,#1}
    \draw[fill=black] (\n) circle[radius=0.75pt, color=black];
\end{tikzpicture}}

\newcommand{\Ngon}[2][black]{\begin{tikzpicture}[node distance=0.25cm,baseline={-0cm-0.5*height("$=$")}]
  \node[regular polygon,regular polygon sides=#2,draw,minimum size=0.5cm](#2-gon){};
  \foreach \X in {1,...,#2}{\draw[fill=#1] (#2-gon.corner \X) circle[radius=1.25pt];}
\end{tikzpicture}}

\newcommand{\sixkite}[1][black]{\begin{tikzpicture}[node distance=0.5cm,baseline={0.15cm-0.5*height("$=$")}]
  \node (0) {};
  \node (1) [right of=0] {};
  \node (2) [right of=1] {};
  \node (3) [above of=2] {};
  \node (4) [left of=3] {};
  \node (5) [above of=0] {};
  \draw (5.center) to (0.center);
  \draw (0.center) to (1.center);
  \draw (1.center) to (2.center);
  \draw (2.center) to (3.center);
  \draw (3.center) to (4.center);
  \draw (4.center) to (1.center);
  \draw[fill=#1] (0) circle[radius=1.25pt, color=black];
  \draw[fill=#1] (1) circle[radius=1.25pt, color=black];
  \draw[fill=#1] (2) circle[radius=1.25pt, color=black];
  \draw[fill=#1] (3) circle[radius=1.25pt, color=black];
  \draw[fill=#1] (4) circle[radius=1.25pt, color=black];
  \draw[fill=#1] (5) circle[radius=1.25pt, color=black];
\end{tikzpicture}}

\newcommand{\lct}[1][black]{\begin{tikzpicture}[node distance=0.5cm,baseline={0.15cm-0.5*height("$=$")}]
  \node (0) {};
  \node (1) [above of=0] {};
  \node (2) [right of=1] {};
  \node (3) [right of=0] {};
  \node (6) [right of=3] {};
  \node (4) [above of=6] {};
  \node (5) [right of=4] {};
  \node (8) [below of=6] {};
  \node (7) [right of=8] {};
  \draw (0.center) to (1.center);
  \draw (0.center) to (3.center);
  \draw (1.center) to (2.center);
  \draw (2.center) to (3.center);
  \draw (3.center) to (6.center);
  \draw (3.center) to (4.center);
  \draw (4.center) to (5.center);
  \draw (4.center) to (6.center);
  \draw (5.center) to (6.center);
  \draw (6.center) to (7.center);
  \draw (6.center) to (8.center);
  \draw[fill=#1] (0) circle[radius=1.25pt, color=black];
  \draw[fill=#1] (1) circle[radius=1.25pt, color=black];
  \draw[fill=#1] (2) circle[radius=1.25pt, color=black];
  \draw[fill=#1] (3) circle[radius=1.25pt, color=black];
  \draw[fill=#1] (4) circle[radius=1.25pt, color=black];
  \draw[fill=#1] (5) circle[radius=1.25pt, color=black];
  \draw[fill=#1] (6) circle[radius=1.25pt, color=black];
  \draw[fill=#1] (7) circle[radius=1.25pt, color=black];
  \draw[fill=#1] (8) circle[radius=1.25pt, color=black];
\end{tikzpicture}}

\newcommand\sii{\sigma_i}
\newcommand\sij{\sigma_j}

\newcommand\mup{\mu_\pi}
\newcommand\sums{\sum_{\{\boldsymbol\sigma\}}}
\newcommand\Z{\mathcal{Z}}

\newcommand\e{\mathrm{e}}

\newcommand\tg{\tilde{\gamma}}
\newcommand\tn{\tilde{n}_\gamma}

\newcommand\ngraph[1]{\ensuremath{n_{\resizebox{!}{0.3\baselineskip}{#1}}}}
\newcommand\tngraph[1]{\ensuremath{\tilde{n}_{\resizebox{!}{0.3\baselineskip}{#1}}}}

\usepackage{amsmath}
\usepackage{xcolor}
\usepackage{natbib}
\begin{document}

\title{Key interaction patterns in proteins revealed by cluster expansion of the partition function}
\author{Matteo Tajana}
\affiliation{Department of Physics, Universit\`a degli Studi di Milano, via Celoria 16, 20133 Milano, Italy}
\author{Antonio Trovato}
\affiliation{Department of Physics and Astronomy ``G.\ Galilei'', Universit\`a degli Studi di Padova and INFN, via Marzolo 8, 35121 Padova, Italy}
\author{Guido Tiana}
\affiliation{Department of Physics and Center for Complexity and Biosystems, Universit\`a degli Studi di Milano and INFN, via Celoria 16, 20133 Milano, Italy}

\date{\today}

\begin{abstract}
The native conformation of structured proteins is stabilized by a complex network of interactions. We analyzed the elementary patterns that constitute such network and ranked them according to their importance in shaping protein sequence design. To achieve this goal, we employed a cluster expansion of the partition function in the space of sequences and evaluated numerically the statistical importance of each cluster. An important feature of this procedure is that it is applied to a dense, finite system. We found that patterns that contribute most to the partition function are cycles with even numbers of nodes, while cliques are typically detrimental. Each cluster also gives a contribute to the sequence entropy, which is a measure of the evolutionary designability of a fold. We compared the entropies  associated with different interaction patterns to their abundances in the native structures of real proteins.
\end{abstract}

\maketitle

\section{Introduction}

The native state of proteins is stabilized by a complex set of heterogeneous interactions among their amino acids. Approximating these interactions as two--body and short--ranged, one can simplify the description of the native state in terms of a network whose nodes are the amino acids and whose edges are their mutual interactions (for a review, see ref.~\cite{DiPaola2013}).

The analysis of the interaction networks between amino acids of single--domain proteins is not trivial due to the fact that they are usually small, and thus the features that characterize quantitatively the associated networks vary on limited ranges. Nonetheless, such networks were found to be usually small--world \cite{Vendruscolo2002} and display clusters of interactions \cite{Kannan1999}, which are often related to the biological properties of the protein.
Even small clusters of contacts can be highly relevant for protein kinetics \cite{Vendruscolo2001}, by nucleating the folding process \cite{Abkevich1994} and for the thermodynamics stabilization of the native state \cite{Tiana2004c}.

To produce well--folding proteins, evolution was supposed to insert in the nodes of the network the twenty types of amino acids with the purpose of obtaining a particularly low energy in the native conformation \cite{Shakhnovich1993}, trying to minimize the frustration of the system \cite{Bryngelson1987}. In this respect, evolution can be regarded as a stochastic process in the space of sequences controlled by the energy of the native conformation, being the rest of the conformational spectrum sequence--independent \cite{Shakhnovich1993a}. Assuming that it reached a stationary state \cite{Rost1997}, evolution can then be described by a canonical ensemble associated to the space of sequences, the temperature $T_s$ being a parameter that sets the evolutionary bias towards sequences having a low energy in the native states. 

It was suggested that some protein conformations are more `designable' than others, being able to act as native state of more sequences \cite{Li1996}.  Several  properties of the native conformation have been associated with an increased designability, including conformational symmetries \cite{Govindarajan1996,Wolynes1996,Maritan2000}, abundance of local interactions \cite{Plaxco1998}, of returning loops \cite{England2003}, of loops of specified size \cite{Berezovsky2000}, or a large local entropy \cite{Negri2021}.

The goal of the present work is to study if there are specific patterns in the interaction network that can favour the evolution of the proteins displaying them. 
In principle, one could study the partition function of the protein in the space of sequences, a problem analogous to a Potts model on an irregular lattice defined by the contact map of the native conformation. However, this approach has two drawbacks. First, it can be used only for small systems because the number of protein sequences of size $N$ to be summed in the partition function scales as $20^N$, a very fast growth even for small--sized proteins. Moreover, calculating the partition function for a given protein gives only information on that protein, while we would like to learn what are the most important `bricks' that drive sequence design in general. 

For this purpose, we employed a standard tool of statistical mechanics, that is the cluster expansion \cite{Huang1987}, on the network of interactions that stabilize the native states of proteins. The terms of the cluster expansions are the `bricks', which are independent on the overall contact map of the protein. 
There are some important differences with the standard cluster expansion of dilute gases. Most importantly, proteins, even large ones are small systems and the thermodynamic limit cannot be applied. Also, proteins are dense systems and we do not expect that the highest--degree terms of the expansion vanish. Our approach is to evaluate what are the most important terms of each degree and eventually to infer what are the elementary structures of the proteins that most contribute to the partition function. On the other hand, studying a finite system we do not need to worry about the convergence of the expansion.

In Sect.~\ref{sect:real} we will compare the elementary structures with highest entropy in the space of sequences with those found most commonly in natural proteins.

\section{The cluster expansion}

The partition function in the space of sequences, for a fixed native conformation, is
\begin{equation}\label{eq:Z}
  \Z = \sum_{\{\boldsymbol\sigma\} } \exp \left[-\beta_s \sum_{i<j}  E_{\sii\sij}\Delta_{ij}\right],
\end{equation}
where $\boldsymbol\sigma$ is the $N$-element vector that specifies a protein sequence, $\beta_s\equiv 1/T_s$, $E_{\sigma\pi}$ is the interaction energy between amino acids of type $\sigma$ and $\pi$, and $\Delta_{ij}$ is the contact map of the protein, that is a binary matrix that specifies what amino acids is in spatial contact with what amino acid in the native conformation. Two amino acids are defined to be in contact if any two atoms belonging, respectively, to each of them have a distance lower than a threshold $d_c$ and they are separated by at least other $i_\text{min}$ residues along the chain.

Defining the Mayer functions $f_{ij}\equiv \e^{-\beta_s E_{\sii \sij}\Delta_{ij}}-1$, the partition function can be rearranged as
\begin{equation}
  \Z =  \sums 1 + \sums \sum_{i<j} f_{ij} + \sums \sum_{i<j, k<l} f_{ij} f_{kl} +\sums \sum_{i<j, k<l, m<n} f_{ij} f_{kl} f_{mn} + \cdots. \label{eq:cluster_expansion}
\end{equation}
Due to the presence of the binary function $\Delta_{ij}$ in the Mayer's function, the different terms of this sum reflect the geometry of the interactions and can be represented as graphs like, for example
\begin{equation}
  \twonodes = \sums \sum_{i<j} f_{ij} = \ngraph{\twonodes} q^{N-2}\sum_{\sigma\pi}  (\e^{-\beta_s E_{\sigma\pi}} -1)= \ngraph{\twonodes} q^N \times  \twonodes[white] 
  \label{eq:edge},
\end{equation}
where $\ngraph{\twonodes}=\sum_{i<j}\Delta_{ij}$ is the number of times that this specific graph appears in the protein, $q$ is the number of types of amino acids and 
\begin{equation}
    \twonodes[white] \equiv \frac{1}{q^2} \sum_{\sigma\pi}\e^{-\beta_s E_{\sigma\pi}}-1
    \label{eq:edgeEmpty}
\end{equation}
 is a shorthand notation for the term containing the average over the types of amino acids. Graphs with empty circles will be called `interaction graphs' because they depend only on the interaction matrix $E_{\sigma\pi}$ and not on the overall structure of the protein.  Similarly, 
\begin{equation}
     \drawpath{3} = \sums \sum_{i<j<k} \left( f_{ij} f_{jk} + f_{ik} f_{jk} + f_{ij} f_{ik}\right)
    = \ngraph{\drawpath{3}} q^{N} \times \drawpath[white]{3},
     \label{eq:2path}
\end{equation}
where 
\begin{equation}
    \drawpath[white]{3}\equiv \frac{1}{q^3} \sum_{\sigma\pi\rho} \e^{-\beta (E_{\sigma\pi} + E_{\pi\rho})}-2\frac{1}{q^2} \sum_{\sigma\pi}\e^{-\beta_s E_{\sigma\pi}}+1,
    \label{eq:2pathEmpty}
\end{equation}
and $\ngraph{\drawpath{3}}=\sum_{i<j<k}\left(\Delta_{ij} \Delta_{jk} + \Delta_{ik} \Delta_{jk} + \Delta_{ij} \Delta_{ik}\right)$, and so on for the other graphs.  Using this formalism, the partition function reads
\begin{equation}
    \Z = q^N + \twonodes + \drawpath{3} + \twonodes\twonodes+ \Ngon{3} + \drawpath{4} + \stargraph{3}{0.3} + \twonodes\drawpath{3} + \twonodes\twonodes\twonodes + \cdots.
    \label{eq:z}
\end{equation}

For proteins, the number of terms is finite and the highest--order term can be, at most, of order of the number of contacts. Anyway, the strategy we will follow is that of selecting {\it a priori} a limited set $\Gamma$ of connected graphs to investigate that are common to realistic protein conformations and study their contribution to the partition function. Vice versa, the summation of all possible graphs that compose the interaction network of a protein is equivalent to summing its partition function, is computationally hard and yields results that can be hardly generalized to other conformations.

There are two properties that are useful in this expansion. First, in each term the `energetic' and the `geometric' parts of the graph factorize, and consequently all of them are in the form 
\begin{equation}
    \gamma=n_\gamma q^N \tg,
    \label{eq:factorization}
\end{equation}
where $\gamma$ denotes the kind of graph, $n_\gamma$ is the number of times that the specific graph appears in the native conformation of the protein and $\tg$ is the corresponding interaction graph, that is
\begin{equation}
    \tg\equiv \frac{1}{q^N}\sums \prod_{(i,j) \in \text{edges}} \left( \e^{-\beta_s E_{\sigma_i\sigma_j}} - 1 \right)
    \label{eq:white}
\end{equation}
Moreover, if the native conformation of a protein displays disjoint interaction patterns, the partition function factorizes into their connected components. 

An unhandy feature of Eq.~(\ref{eq:z}) is that it contains not only connected graphs describing the interaction patterns contained in the native structure, but also all possible disjoint combinations of them. If the protein were an infinite system, with all residue pairs in contact with each other, one could use the connected--graph theorem \cite{Huang1987} and the partition function in the grand-canonical ensemble could be written as the exponential of the sum of contributions from connected graphs only. Since we are focusing on finite--size properties, we followed another strategy.

The value of a disjoint interaction graph is just the product of its components. The problem is to count the number of occurrences of the corresponding pattern in the protein. For example, 
\begin{equation}
    \twonodes\twonodes=q^N \left[ \binom{\ngraph{\twonodes}}{2}-\ngraph{\drawpath{3}} \right]\cdot\twonodes[white]^2
    \label{eq:twoNodes}
\end{equation}
where the binomial coefficient counts the number of ways one can extract two links from the native contact map with $\ngraph{\twonodes}$ edges and $\ngraph{\drawpath{3}}$ removes from that count the number of instances in which the two links are sharing a common node, thus not contributing to Eq.~(\ref{eq:twoNodes}). 
Similarly,
\begin{equation}
    \twonodes\drawpath{3} = q^N \left[ \binom{\ngraph{\twonodes}}{1}\binom{\ngraph{\drawpath{3}}}{1}-\ngraph{\drawpath{4}}-\ngraph{\stargraph{3}{0.3}}-\ngraph{\Ngon{3}}-\ngraph{\drawpath{3}}\ngraph{\twonodes\drawpath{3}\Big{/}\drawpath{3}}\right]  \cdot\twonodes[white]\drawpath[white]{3},
    \label{eq:linkAngle}
\end{equation}
where the negative terms enumerate all possible cases in which the connected components $\twonodes$ and $\drawpath{3}$ can be found as superimposing elements in the native contact map, that is sharing common nodes and/or edges. In particular, $\ngraph{\twonodes\drawpath{3} \Big{/} \drawpath{3}}=2$ counts in how many ways the disjoint components of the graph $\twonodes\drawpath{3}$ can be superimposed to form the graph $\drawpath{3}$.
As a matter of fact, all terms of Eq.~(\ref{eq:z}) can be written in a form analogous to Eq.~(\ref{eq:twoNodes}),
\begin{equation}
    \Z=q^N\sum_{k_1=0}^{\ngraph{\twonodes}}\sum_{k_2=0}^{\ngraph{\drawpath{3}}}\cdots \left[\binom{\ngraph{\twonodes}}{k_1}\binom{\ngraph{\drawpath{3}}}{k_2}\cdots-\nu_{k_1,k_2,...}\right]\cdot\twonodes[white]^{k_1}\drawpath[white]{3}^{k_2}\cdots,
    \label{eq:withNu}
\end{equation}
where a given choice of ${k_1,k_2,...}$
corresponds to an in general disjoint graph, and $\nu_{k_1,k_2,...}$ is the number of ways this graph can be formed by extracting its disjoint components from the native pattern of interactions in such a way that at least some of those components superimpose with each other.
For example, $\nu_{2,0,0,\dots}=\ngraph{\drawpath{3}}$, $\nu_{1,1,0,\dots}=\ngraph{\drawpath{4}}+\ngraph{\Ngon{3}}+\ngraph{\stargraph{3}{0.3}}+\ngraph{\drawpath{3}}\ngraph{\twonodes\drawpath{3}\Big{/}\drawpath{3}}$ (while $\nu_{0,0,0,\dots}=0$ for the graph with no links, and $\nu_{1,0,\dots}=\nu_{0,\dots,0,1,0,\dots}=0$ for all connected graphs).

Importantly, the sums in Eq.~(\ref{eq:withNu}) are taken only for connected graphs in our preselected set $\Gamma$; otherwise, Eq.~(\ref{eq:withNu}) would be as intractable and ungeneralizable as Eq.~(\ref{eq:Z}). 
Note also that in general several prefactors $[\dots]$ in Eq.~(\ref{eq:withNu}) can be zero. 

One can evaluate $\Z$ in mean--field theory by applying an approximation similar to saddle--point to the sums in Eq.~(\ref{eq:withNu}).
The largest term in each binomial sum is that associated with $k_\gamma=n_\gamma\tg/(\tg+1)$. If $\tg\gg 1$ then the dominant term is $k_\gamma\approx n_\gamma$, corresponding to the disjoint graphs with the maximum number of connected subgraphs belonging to $\Gamma$. For this graph, the term $\nu$ cannot be neglected because the probability of superimposing nodes is also maximum. The dominant term is then given keeping the subgraphs associated with the largest $\tg$ whose nodes do not overlap. We call $\Gamma^*$ this set and $\tn$ the number of subgraphs $\gamma$ present in the dominant term (in general $\tn < n_\gamma$ due to the no-overlap constraint). In this case the prefactor of the subgraphs in Eq.~(\ref{eq:withNu}) is of the order of 1 and thus
\begin{equation}
    \Z\approx q^N\prod_{\gamma\in\Gamma^*} \tg^{\tn}.
    \label{eq:zProd}
\end{equation}
Note that $\Gamma^*$ and $\tn$ depend on the given native state contact map.
A simple way to calculate them is to order the $\tg\in\Gamma$ from the largest, identifying in this descending order which of them are present in the contact map of the protein, and exclude those whose nodes overlap with the selected ones. 
For example, assuming that
\begin{equation}
    \sixkite[white]>\drawpath[white]{3}>\dots>\twonodes[white]
    \label{eq:order}
\end{equation}
belong to $\Gamma$, one obtains
\begin{equation}
    \Z \left(\lct\right) \approx q^N \times \sixkite[white] \times \drawpath[white]{3}
\end{equation}

A way equivalent to that of Eq.~(\ref{eq:zProd}) to estimate the importance of the graphs is to consider the corresponding free energy, that in the case $\tg\gg 1$ reads
\begin{equation}
    -T_s\log\Z = -T_s \, N\log q -T_s \sum_{\gamma\in\Gamma^*} \tn\log\tg.
    \label{eq:f}
\end{equation}
Note that Eq.~(\ref{eq:f}) defines an approximate free energy, due to using only connected graphs from $\Gamma$ in the cluster expansion and to the saddle point approximation discussed above.

The quantities one needs to study numerically are then the $\log\tg$ for the different types of connected graphs in $\Gamma$.


\section{Numerical evaluation of the contribution of connected graphs}
\label{sect:canonical}

The partition function depends not only on the native contact map, but also on the interaction matrix $E_{\sigma\pi}$ and on the evolutionary temperature $T_s$. 

\subsection{The system}

As interaction matrix we employed that of Table VI of the Miyazawa--Jernigan (MJ) work \cite{Miyazawa1985}, obtained from the statistics of contacts in the Protein Data Bank. 
Although these potentials are a crude simplification of the actual interactions between amino acids \cite{Thomas1996}, they contain the basic features that are needed in a coarse--grained description of proteins as that given by Eq.~(\ref{eq:Z}), like attraction between opposite charges and among hydrophobic residues, repulsion of similar charges, etc.
We have modified the MJ matrix removing the term associated with the interaction between cysteines (substituting it with the matrix average) because disulphide bonds display physical mechanisms, different from those of the other pairs, whose consideraion goes beyond the scope of the present work.  As controls, we used random matrices in which the elements of the MJ matrix are reshuffled, loosing the correlations present in the original MJ, and random matrices with the same Gaussian distribution of elements as the original MJ ($\overline{E}=0.02$ and $\sigma_E=0.30$). Moreover, we also tested the effect of random matrices, either reshuffled from the MJ or extracted from a Gaussian distribution, in which all the diagonal elements are set to $\overline{E}$.

\begin{figure}
    \centering
    \includegraphics[width=10cm]{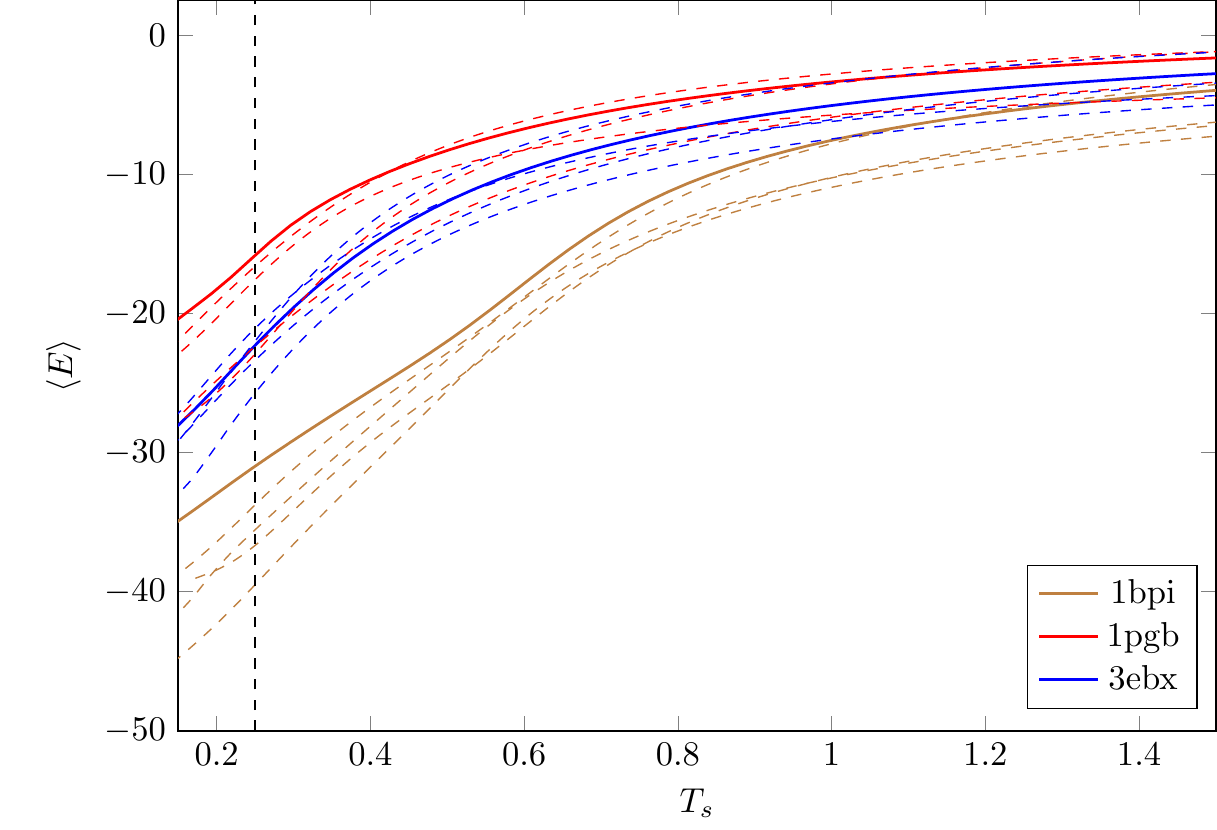}
    \caption{The average energy $\langle E\rangle$ as a function of evolutionary temperature $T_s$ for three proteins (in the legenda, their pdb codes) interacting with the MJ interaction matrix (solid curves) and for randomly--reshuffled matrices (dashed curves). The vertical bar indicates the temperature we chose, that is in the low--temperature regime for all systems.}
    \label{fig:av_energy}
\end{figure}

To set a realistic value for $T_s$ we explored the thermodynamic properties of the sequence space of three small proteins (the villin headpiece [pdb code: 1bpi] of 36 residues, protein G [1pgb] of 56 residues, erabutoxin B [3ebx] of 62 residues) in the canonical ensemble at varying evolutionary temperatures with an adaptive simulated tempering algorithm \cite{Tiana2011}. 
We calculated numerically (Fig.~\ref{fig:av_energy}) the average energy $\langle E \rangle$ of the protein as a function of $T_s$ for the three proteins interacting with the MJ matrix and with randomly--reshuffled matrices. As expected \cite{Ramanathan1994}, at low temperatures, $\langle E \rangle$ is a linear function of $T_s$, while at high temperatures $\langle E \rangle\sim 1/T_s$ which is typical of the random energy model. The crossover between the two behaviors is slightly system--dependent and takes place between $T_s^{c1}=0.5$ and $=1.0$. The statistical properties of natural sequences are compatible with the low--temperature regime \cite{Franco2019}; we then focused our study on $T_s=0.25$, at which all curves are in the linear regime.

\subsection{Contribution of the connected graphs} \label{sect:graphs}

Due to the factorization of all graphs $\gamma$ into a geometric factor $n_\gamma$ and an interaction term $\tg$, it is interesting to identify the largest values $\tg$ that yield the main contributions to the partition function. They can be either positive or negative (cf.\ Eqs.~(\ref{eq:edgeEmpty}) and (\ref{eq:2pathEmpty})) according to whether the typical contribution is attractive or repulsive, respectively. In principle, an even number of negative terms can result in a large positive contribution to the partition function (\ref{eq:withNu}). However, this would be a stabilization mechanism hardly to be generalized. Therefore, in our cluster expansion strategy we will consider only connected graphs with a large {\em positive} contribution as possible subsets of $\Gamma^*$.

\begin{figure*}
    \centering
    \includegraphics[width=\linewidth]{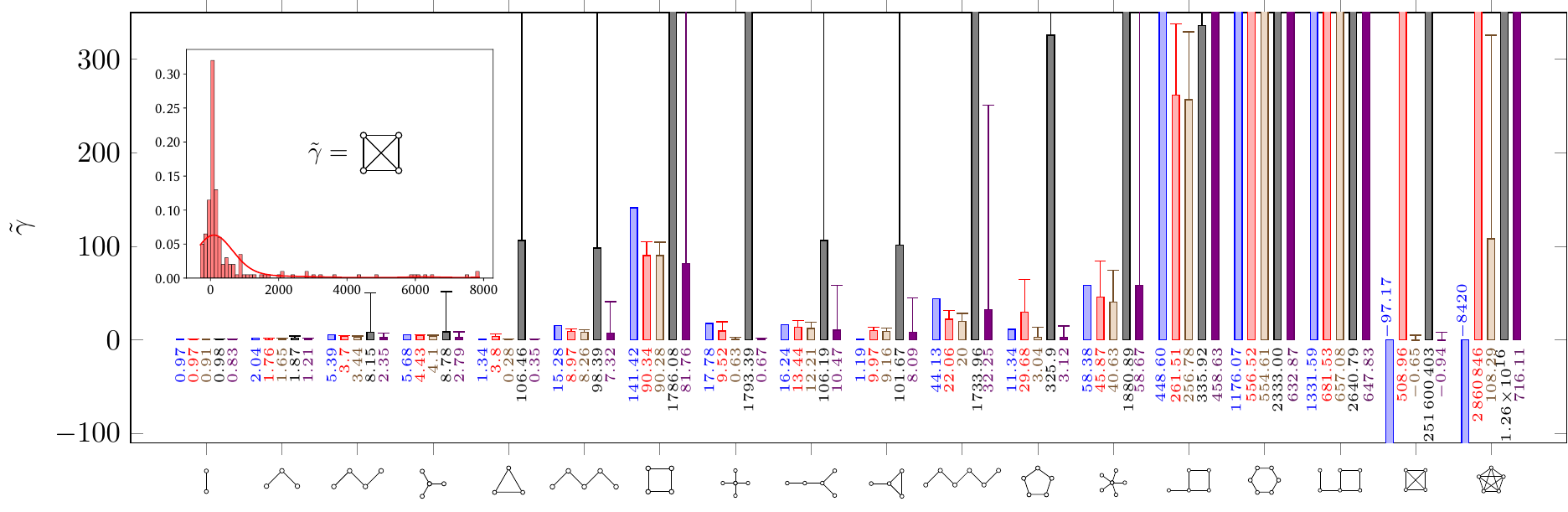}
    \caption{The values of $\tg$ for graphs interacting with the modified MJ matrix (blue bars), with  randomly-reshuffled matrices (red bars), with randomly reshuffled matrices setting all diagonal elements equal to $\overline{E}$ (brown bars), with random matrices with average $\overline{E}$ and standard deviation $\sigma_E$ (gray bars) and random matrices setting all diagonal elements equal to $\overline{E}$ (purple bars). For random matrices is indicated the average over 100 realizations and the thin bars indicates the standard deviation. In the inset, the distribution of values obtained for the 4--link clique interacting with the reshuffled matrices.}
    \label{fig:canonical}
\end{figure*}

The values for all connected graphs with up to 4 links and for some graphs with a larger number of links, computed for the MJ matrix, (blue bars in Fig.~\ref{fig:canonical}) are compared with the average contributions of random matrices. If the native conformation is made of $\ngraph{\twonodes}$ independent contacts, the approximate free energy (\ref{eq:f}) is straightforwardly proportional to the logarithm of $\twonodes[white]=0.97$ multiplied by $\ngraph{\twonodes}$. If these contacts are assembled in more complex patterns, this value of the free energy is corrected by including the corresponding graphs in $\Gamma^*$.

Note that, in the simple ``diluted'' example of $\ngraph{\twonodes}$ independent contacts, one can explicitly sum the binomials in Eq.~(\ref{eq:withNu}), so that the {\em exact} result is $\log\Z = N\log q + \ngraph{\twonodes}\log(\twonodes[white]+1)$, as it should. The cluster expansion strategy that we propose fails in this simple example, but is in fact meant for dense connected graphs.

\begin{figure}
    \centering
    \includegraphics[width=8cm]{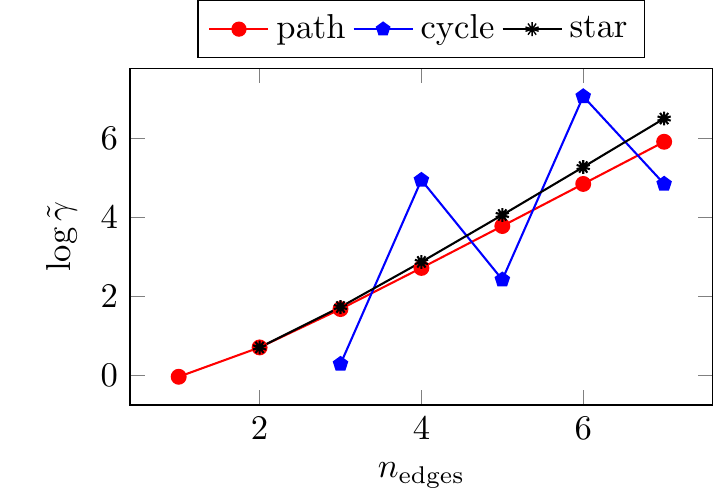}
    \caption{The same data for the modified MJ matrix, shown as blue bars in Fig.~\protect\ref{fig:canonical} for paths, cycles and stars, displayed in semi--logarithmic scale as a function of the number of edges.}
    \label{fig:graphs}
\end{figure}

Since the protein is a dense system, the value of $\tg$ tends to increase with the number of links. The graphs displaying the largest relative contribution to the partition function for the MJ matrix are the cycles with even number of links (Fig.~\ref{fig:graphs}). The largest contribution to this graphs comes from arrangements of pairs of residues of the kind $\alpha\beta\alpha\beta\dots$, where the $E_{\alpha\beta}$ belong to the low--energy tail of the distribution of energies. These arrangements are equally likely for the random matrices; however, for purely random matrices whose elements are extracted from the Gaussian
distribution with the same first two
moments as the MJ interactions, these
arrangements remain the favoured ones
only when constraining the diagonal matrix terms (see discussion below).
On the other hand, the reshuffled MJ matrices display similar values for the even cycles, both with and without the constraint on diagonal terms.

The fact that the contributions of even cycles with the MJ matrix are larger than those with the randomly reshuffled  matrices indicates that the MJ matrix displays specific correlations between residues, like the case of charged and of hydrophobic residues.

On the other hand, cycles with an odd number of links cannot benefit from simply looping the elements with low $E_{\alpha\beta}$, but their value is controlled by the existence (by chance or by specific correlations in the interaction matrix) of a residue kind $\gamma$ such that also $E_{\alpha\gamma}$ and $E_{\beta\gamma}$ are attractive enough. The fact that the contributions for odd cycles (triangles and pentagons) interacting with the reshuffled matrix, with the constraint on diagonal terms, are smaller than those interacting with the MJ matrix suggests the presence of anti--correlations in the interactions between natural amino acids (i.e.,\ if $E_{\alpha\beta}\ll 0$ then the same is not true for $E_{\alpha\gamma}+E_{\beta\gamma}$).

Also the values for paths (i.e.,\ unlooped chains of linked nodes) and stars (i.e.,\ nodes linked only to a central node) yield a positive, non--negligible contribution which is similar between paths and stars with the same number of links. In all cases the graphs associated with the MJ matrix are larger than the randomly--reshuffled ones. 

The different control models interacting with random energies display specific features (Fig.~\ref{fig:canonical}). The graphs calculated with matrices randomly reshuffled from the MJ do not change appreciably their behavior whether the diagonal elements are randomly reshuffled like all the other elements (red bars) or they are set equal to a constant value equal to the average $\overline{E}$ of the matrix. On the contrary, in the case of matrices obtained by a random extraction of energies from a Gaussian distribution (gray bars), one can obtain average values that are very large and not self--averaging (cf.\ the error bars), corresponding to the realizations of the matrix in which a particularly negative element appears on the diagonal. In this case, the value of all graphs become much larger then in the typical case because of the multiple appearance of the corresponding amino acids. In fact, matrices in which off--diagonal elements are generated randomly and diagonal elements are set to $\overline{E}$ (cf.\ purple bars) do not display this effect. 

The overall scaling of the different free contributions with the number of links (Fig.~\ref{fig:graphs}), suggests that a sensible way of identifying $\Gamma^*$ in the contact map of a protein is first to search for even cycles, then for stars and paths.

Cliques (i.e.,\ fully--connected sets of nodes) made of four or more nodes yield a negative contribution (cf.\ Fig.~\ref{fig:canonical}). This appears as a consequence of the frustration of the system; according to the MJ matrix it is not possible to allocate more than three amino acids in such a way that all pairs are attractive. Cliques interacting with random matrices without a constrain on diagonal terms (red and gray bars in Fig.~\ref{fig:canonical}) display large positive averages, but they are not self--averaging. In fact, the average arises from a distribution with a long tail (cf.\ e.g.\ the inset of Fig.~\ref{fig:canonical}), producing a standard deviation that is similar to the average. The reason is that there is a non--negligible probability that some realizations of the random matrix display blocks of negative elements (e.g.,\ a $2\times 2$ block with $E_{\alpha\beta}$, $E_{\alpha\alpha}$, $E_{\beta\beta}\ll -T_s$). The larger the block, the (exponentially) larger the contribution to the clique but the lower the probability that the specific realization of the matrix can achieve such large values. This results in a wide distribution of values for cliques and thus in its non--self--averaging character. Diagonal elements are essential for this mechanism, as shown by the fact that random matrices with `typical' elements $\overline{E}$ on the diagonal cannot display such large values for cliques. 
Cliques are the graphs that can build the largest number of links with a given number of nodes and thus can occasionally produce very large values. 

\subsection{Dependence on the temperature}

\begin{figure}
    \centering
    \includegraphics[width=10cm]{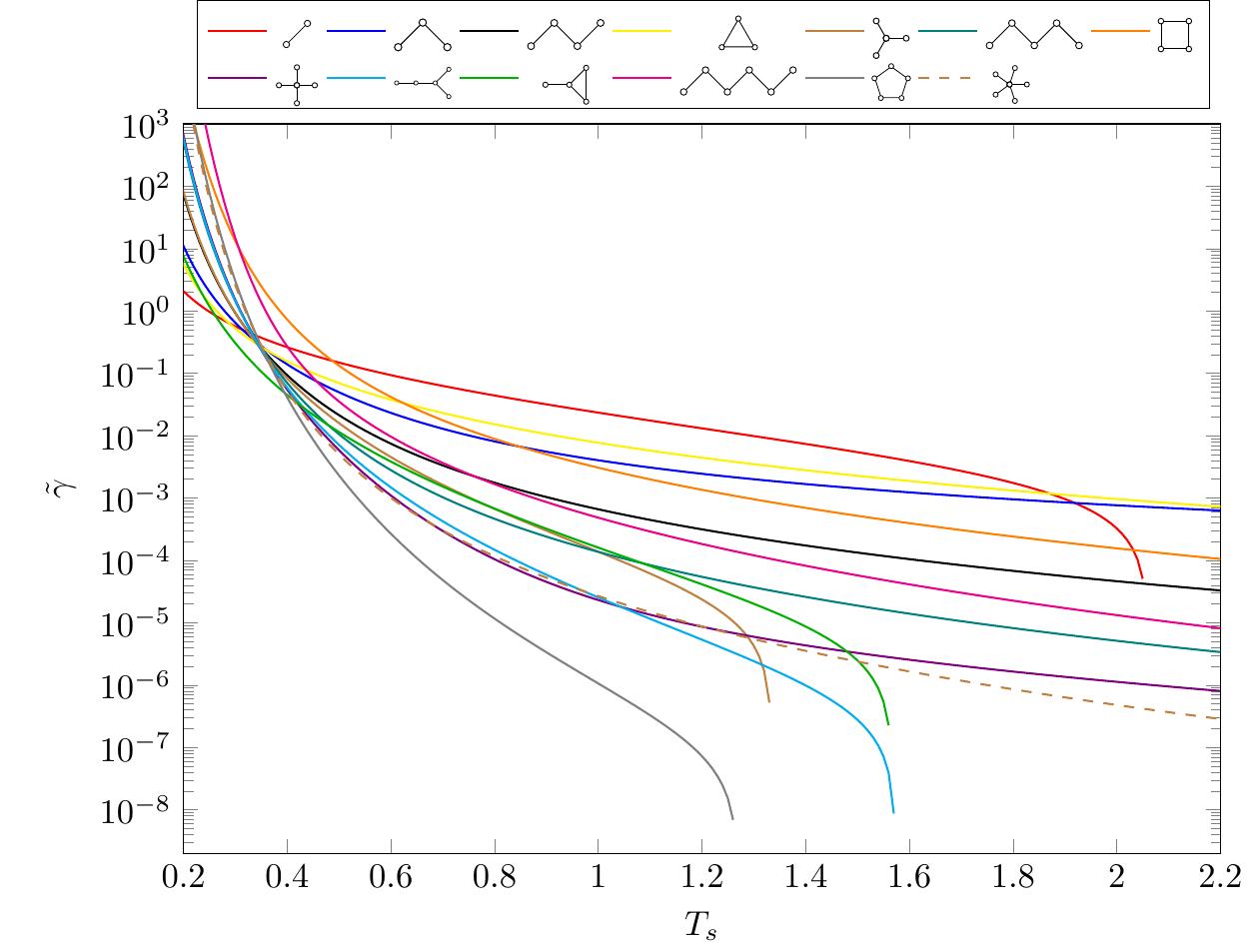}
    \caption{The dependence of some graph cluster expansion contributions with MJ interactions on the evolutionary temperature $T_s$ in semi--log scale. Some curves are displayed only partially because they become negative.}
    \label{fig:t}
\end{figure}

The relative weight of different graphs changes with $T_s$ in a non--straightforward way (Fig.~\ref{fig:t}). Increasing the temperature from $T_s=0.25$ one can observe several crossings between the curves associated with the different graphs. Above $T_s^{c1}\approx 0.6$, when the statistical properties of sequences space are described by the random energy model, the term associated with a single link becomes dominant for a large range of temperatures and, in general, simple small graphs become larger than more complex ones.

All graph contributions go to zero at large evolutionary temperature ($\beta_s\to 0$) because each Mayer function vanishes. Keeping only the first--order term in the high--temperature expansion of each factor of Eq.~(\ref{eq:white}), one obtains
\begin{equation}
    \lim_{\beta\to 0}\tg=\beta^k \frac{(-1)^k}{q^N}\sums E_{\sigma_1\sigma_2}E_{\sigma_3\sigma_4}\cdots E_{\sigma_{2k-1}\sigma_{2k}},
\end{equation}
where the relations among the indexes of the elements of the vector $\mathbf \sigma$ reflect the structure of the graph and $k$ is the number of edges. Thus, one can obtain the $k$--point correlation function of the MJ matrix from the high--temperature limit of the graphs.

This correlation function for the MJ matrix can be either positive or negative. Its sign determines whether $\tg$ approaches zero from the positive or the negative side. For example, in the MJ matrix the two--point correlation function $\drawpath{3}$ is positive while the three--point correlation function $\drawpath{4}$ is negative.

\subsection{Average Energy and Entropy}

Since $-T_s\log\tg$ is the contribution of the different graphs of $\Gamma^*$ to the free energy of sequences, one can split them into an energetic and an entropic contribution. The average energy is $\langle E\rangle=-\partial \log\Z/\partial\beta_s$, and thus the contributions to it of the different graphs are $-\partial \log\tg/\partial\beta_s$.

\begin{figure}
    \centering
    \includegraphics[width=12cm]{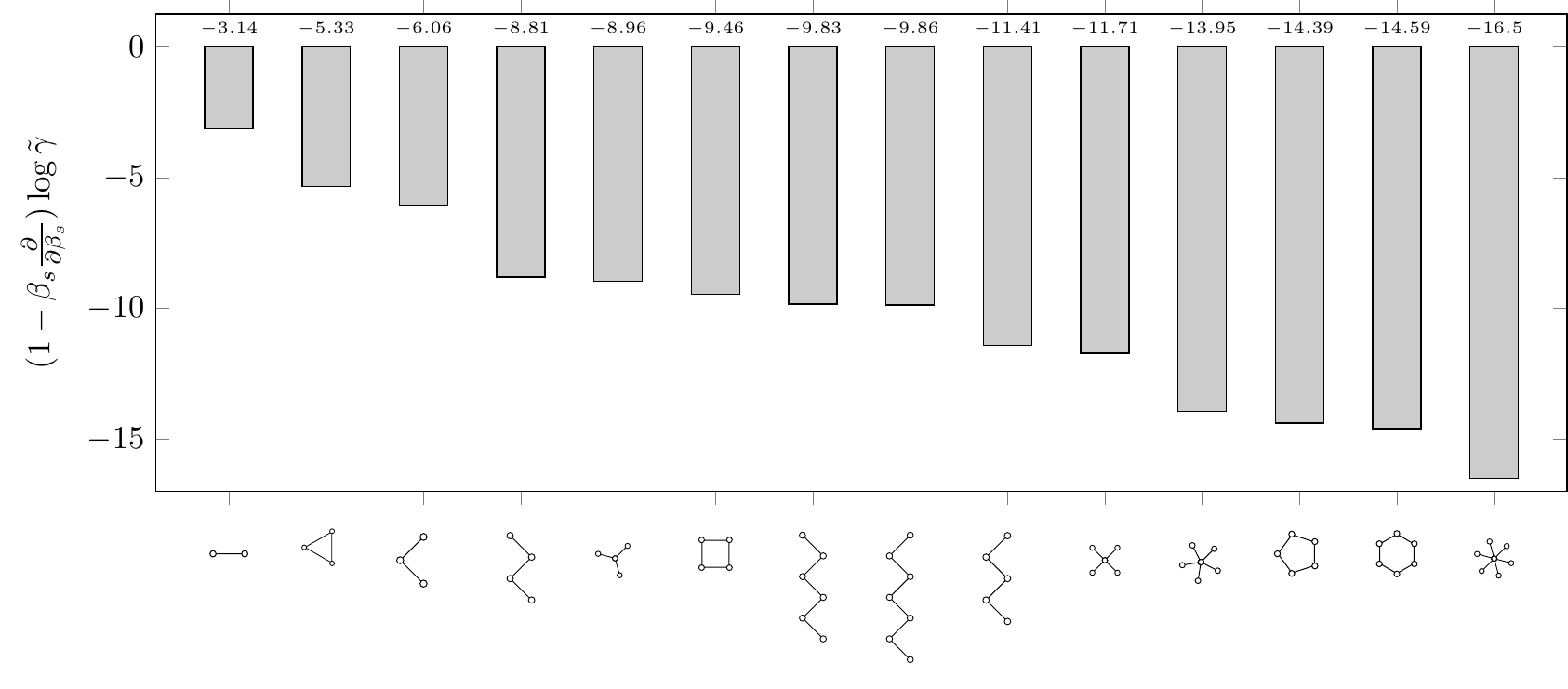}
    \includegraphics[width=12cm]{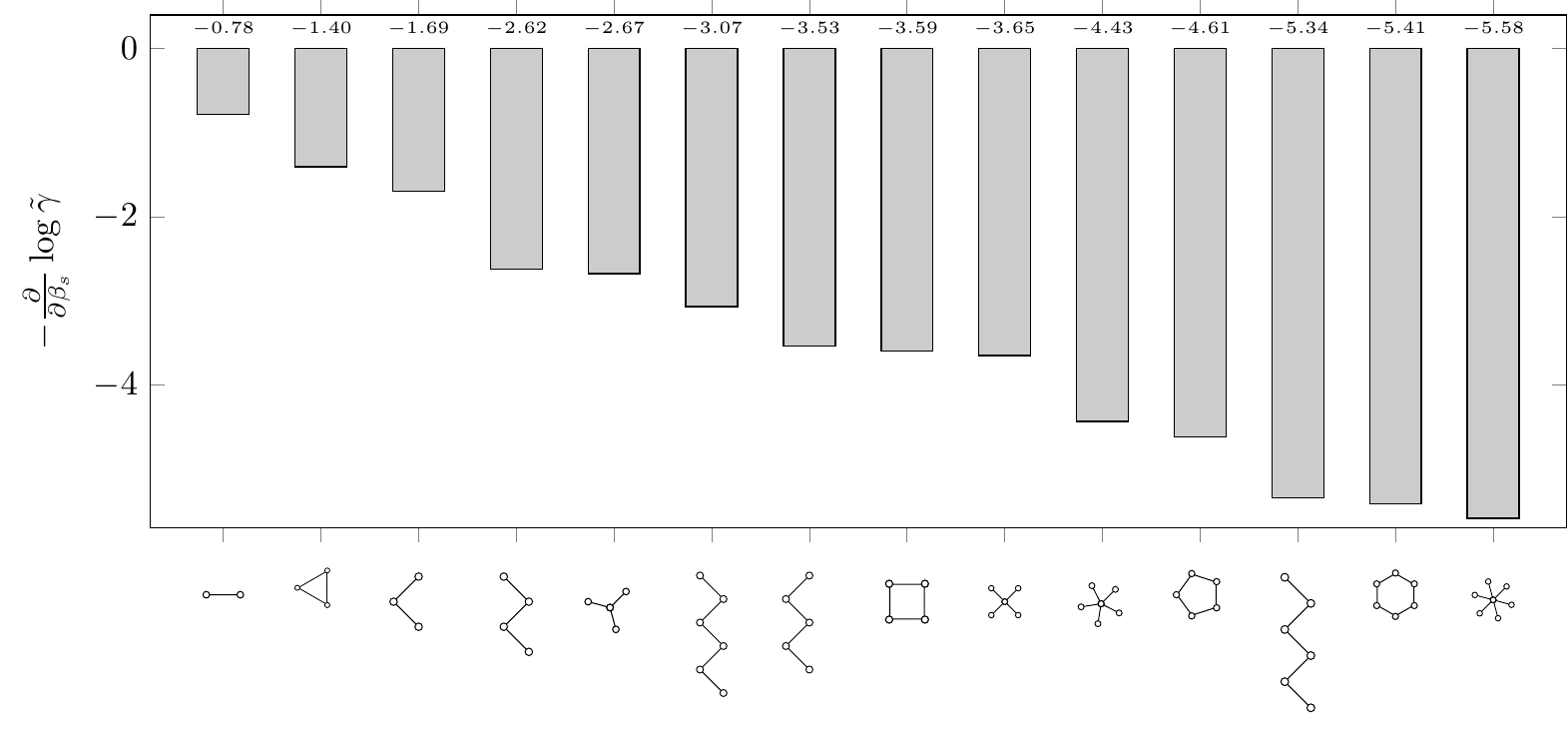}
    \caption{The cluster expansion contribution of different graphs to the entropy in sequence space (upper panel) and to the energy (lower panel) for the MJ matrix at $T_s=0.25$.}
    \label{fig:entropy}
\end{figure}

The contribution of the different graphs belonging to $\Gamma^*$ to the entropy can be obtained applying the thermodynamic relation $S=\beta_s \langle E\rangle+\log\Z$ to Eq.~(\ref{eq:f}), thus obtaining
\begin{equation}
    S = N\log q + {\tngraph{\twonodes}}\left(1-\beta_s\frac{\partial}{\partial\beta_s}\right)\log\left[\twonodes[white]\right] +  {\tngraph{\drawpath{3}}} \left(1-\beta_s\frac{\partial}{\partial\beta_s}\right)\log\left[\drawpath[white]{3}\right] + \cdots.
    \label{eq:s}
\end{equation}
Each term of the sum should be read as an entropic cost with respect to the infinite--temperature entropy $N\log q$ (cf.\ Fig.~\ref{fig:entropy}, where the graphs are ordered according to their contribution to the free energy). Importantly, the entropic contribution of graphs is comparable to their energetic contribution, thus the sequence probability results from a non--trivial balance between stabilization of a given pattern and its degeneracy in sequence space. The contributions from different graphs to energy, entropy and free energy are roughly proportional to each other. Notable exceptions to this pattern are the squares and the 6--paths that yield a energy contribution comparable to the other graphs with the same number of edges but with a lower entropic cost.

\section{Patterns in small systems}

In the case of small networks of few amino acids, the approximations that lead to Eq.~(\ref{eq:f}), the main result of our cluster expansion strategy, would not be justified. In this case, the partition function can be summed directly and the energy and entropy can be calculated exactly as $\langle E\rangle = - \partial\log\Z/\partial\beta$ and $S=\partial (T\log \Z)/\partial T$, respectively. In this context we will consider densities dividing extensive quantities by the number of nodes. In case of disjoint patterns the partition function is the product of the connected patterns and the free energy is the sum of the associated free energies.

\begin{figure}
    \centering
    \includegraphics[width=\linewidth]{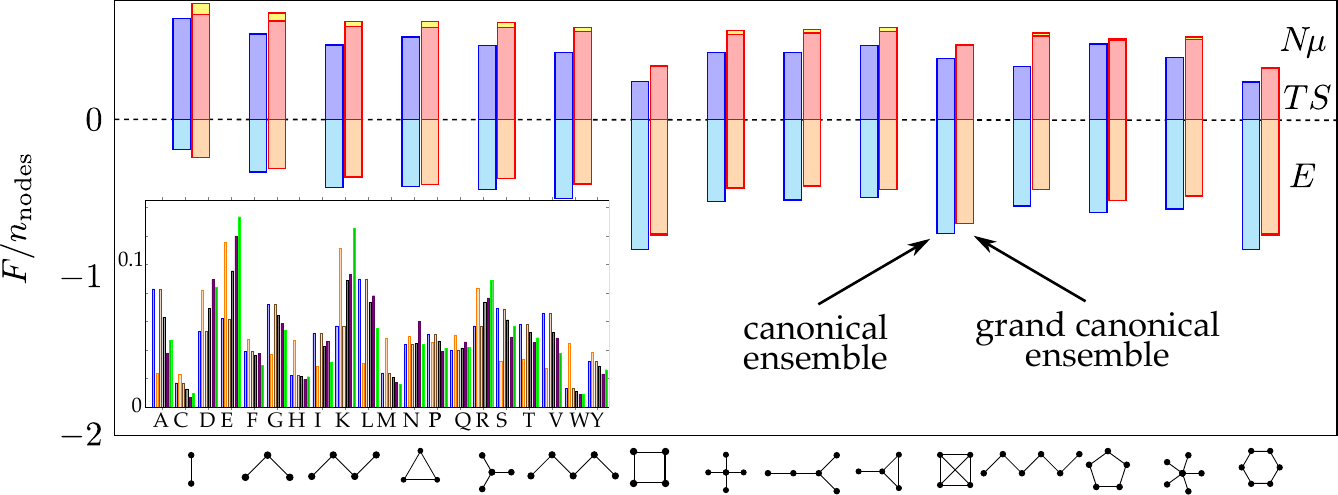}
    \caption{The density of energy (cyan) and entropy (blue) calculated for some small patterns of interaction in the canonical ensemble. The density of energy (orange), entropy (red) and chemical potential (yellow) in the grand--canonical ensemble. In the inset, the natural fraction of the twenty types of amino acids (blue bars), the averages calculated in the high--temperature approximation from the grand--canonical partition function (orange) and those obtained from the iterative estimation of the chemical potentials (brown for an isolated contact, gray for a 2-edge path, purple for a 3-node clique and green for a 3-edge star.}
    \label{fig:grandcan}
\end{figure}

The densities of energy and entropy calculated for some small systems are displayed with cyan and blue bars, respectively, in Fig.~\ref{fig:grandcan}. 
The sum of the two bars gives, for each system, the density of free energy. The largest density of free energy is associated, also in the case of small systems, to even cycles. On the other hand, cliques are now not as penalized as in the case of larger systems. 

A problem associated with this calculation is that the partition function includes, and may be dominated by, sequences with unrealistic concentration of the twenty amino acids. Specifically, the dependence of the stability of proteins on the native energy only \cite{Shakhnovich1993a} requires fixed concentrations. When studying the role of small graphs in a large protein, this is expected to be a minor problem, because the amino acids surrounding the graph of interest act as a reservoir of amino acids. For small systems this can become a problem.

We then studied small systems with different interaction patterns, by adding twenty chemical potentials $\mu_\sigma$ to set the average concentration of amino acids to their natural values \cite{T.E.Creighton1992}.
The associated partition function is
\begin{equation}
    Z = \sum_{\boldsymbol\sigma}\exp\left[-\beta_s\left(\sum_{i<j}E_{\sigma_i\sigma_j}\Delta_{ij}-\sum_i\mu_{\sigma_i}n_{\sigma_i}\right)\right],
\end{equation}
where $n_\sigma$ is the number of residues of kind $\sigma$ in the sequence $\boldsymbol\sigma$. 

To obtain explicit equations for the chemical potentials, one can approximate the average number of amino acids for a given system with that associated with the same number of independent contacts. One can begin evaluating $\mu_\sigma$ in the high--temperature limit, that gives
\begin{equation}
    \mu_\sigma=\frac{1}{\scriptstyle q(2q-1)-1}\left[ \sum_{\rho\pi} E_{\rho\pi} - \sum_\pi E_{\sigma\pi} - \frac{1}{\beta} \log\left(\frac{2n_\sigma}{N}\right) \right].
\end{equation}
However, the average concentrations $\langle n_\sigma\rangle=\beta_s^{-1}\partial\log Z/\partial\mu_\sigma$ calculated in the high--temperature approximations are not in good agreement with the correct ones (cf.\ red bars in the inset of Fig.~\ref{fig:grandcan}) and we corrected them in an iterative way applying 100 times
\begin{equation}
   \mu_\sigma = \frac{1}{\beta} \log \left( p_\sigma \dfrac{\sum_{\rho\pi} \e^{-\beta (E_{\rho\pi} - \mu_\rho - \mup)}}{\sum_\pi \e^{-\beta (E_{\sigma\pi} - \mu_\pi)}} \right) .
\end{equation}
In this way, the average concentrations are well estimated even for non--trivial interaction patterns (cf.\ the inset of Fig.~\ref{fig:grandcan}). Not unexpectedly, in the grand--canonical ensemble the potential energies are slightly less negative than in the canonical ensemble (with the exception of the single link) and the entropies slightly more positive. In fact, the constraints in the composition of the protein eliminate states where few types of amino acids are repeated in a regular way.
Overall, the free energies in the grand--canonical ensemble do not depart drastically from the canonical ones (cf.\ Fig.~\ref{fig:grandcan}).

\section{Analysis of patterns in real proteins} \label{sect:real}

Since the entropy (Eq.~\ref{eq:s}) is a measure of the abundance of protein sequences displaying specific conformational patterns, we analyzed the contact maps of a set of natural proteins and compared the number of occurrences of the patterns (Fig.~\ref{fig:entropy}) with the corresponding entropy. We calculated the contact map of a set of 565 non--redundant proteins obtained from the pdb ($<25$\% sequence similarity).

\begin{figure}
    \centering
    (a)\hfill $\vcenter{\hbox{\includegraphics[height=5.25cm]{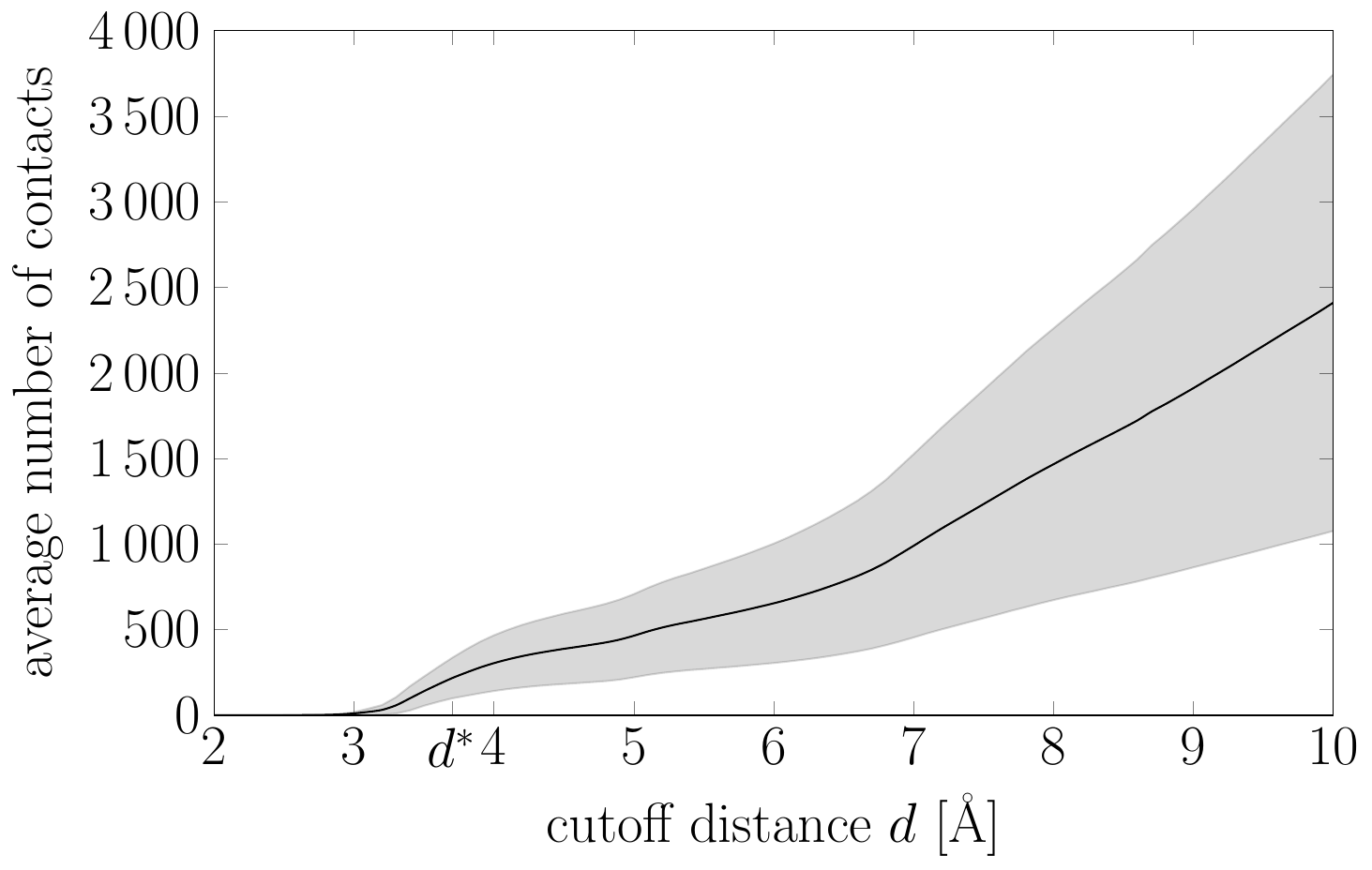}}}$ \hfill
    (b) \quad $\vcenter{\hbox{ \includegraphics[height=4.5cm]{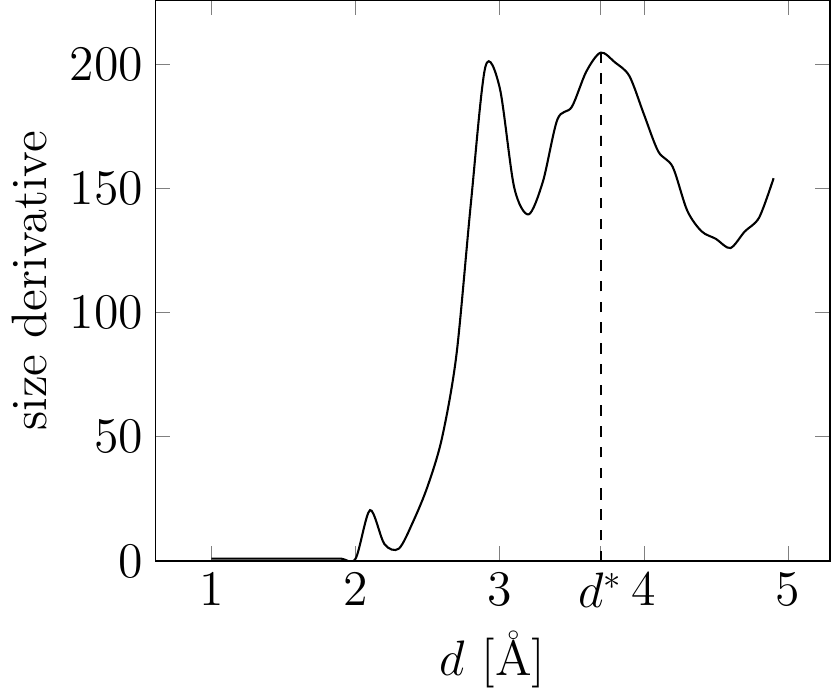}}}$ \hfill \\[0.01cm]
    (c) $\vcenter{\hbox{\includegraphics[width=0.925\linewidth]{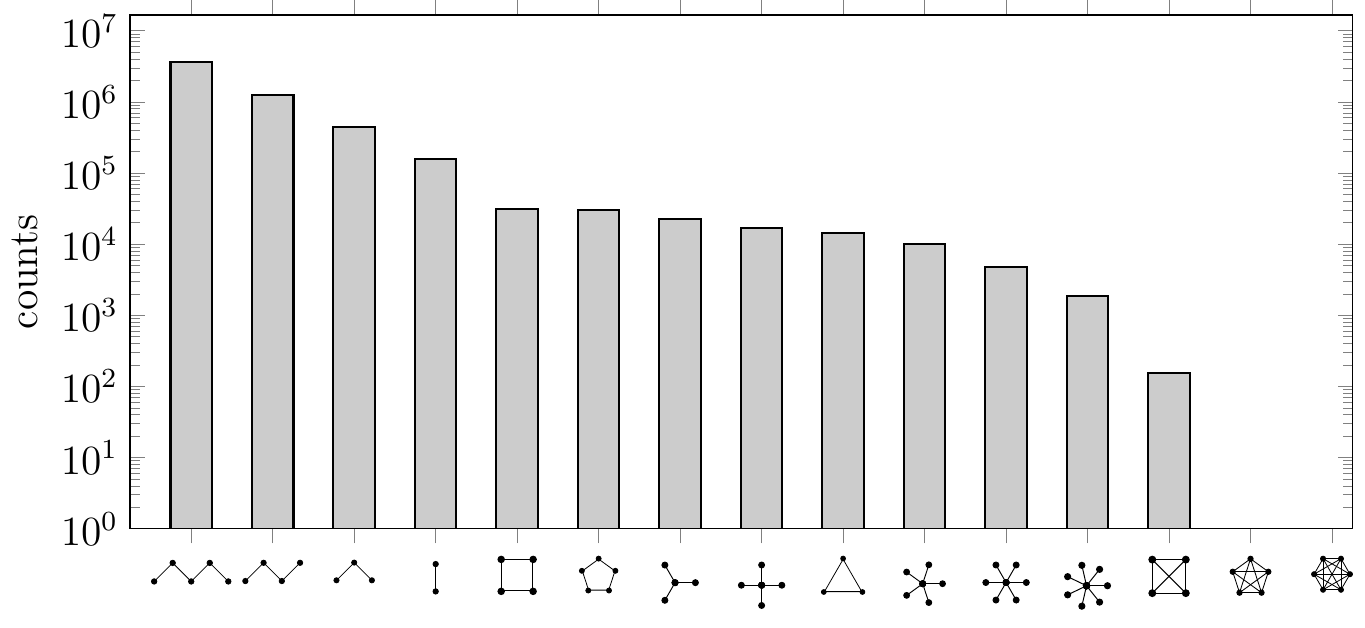}}}$ \hfill \\[0.2cm]
    (d) $\vcenter{\hbox{\includegraphics[width=0.925\linewidth]{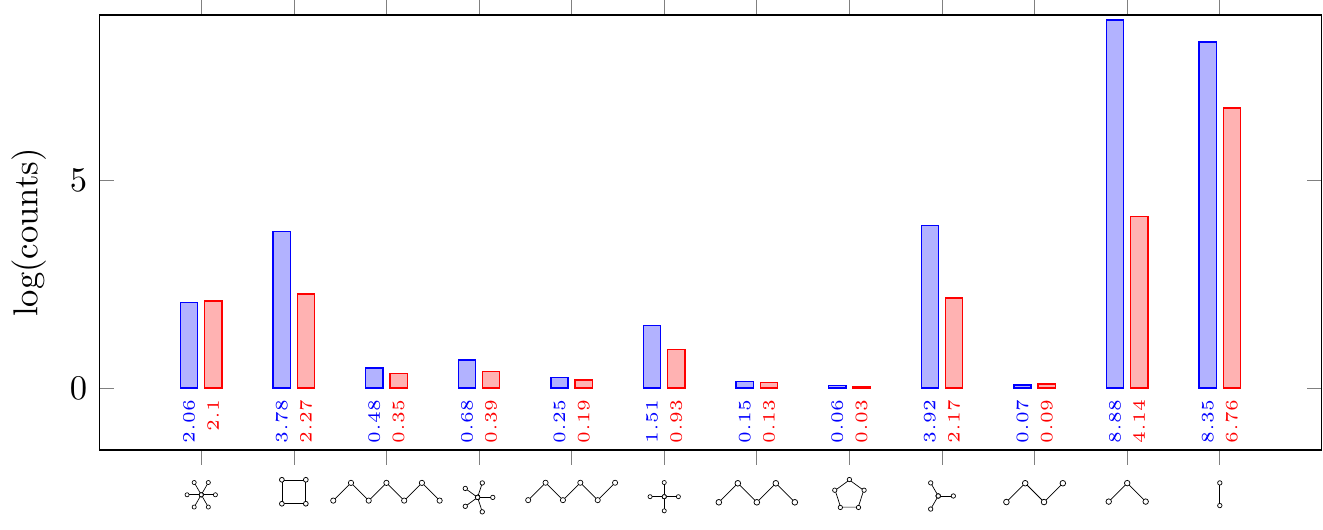}}}$ \hfill \\[0.2cm]
    \caption{(a) Average number of contacts in the protein dataset as a function of the threshold distance $d^*$. (b) Numerical derivative of the size of the graph giant component as a function of $d^*$. (c) The number of graphs found in a non--redundant dataset of natural proteins, (d) the average number of graphs per protein (blue bars), enumerated without double--counting nodes in the order displayed along the horizontal axis from left to right. In red, the same quantity calculated on a set of random decoys.}
    \label{fig:proteins}
\end{figure}

To define the contact map, one has first to define what a contact is. We define two amino acids to be in contact if any two atoms belonging to each of them, respectively, are closer than a distance $d^*$ and farther than three residues along the sequence. We have chosen for $d^*$ the value $3.7${\AA}, at which the mean number of contacts in the dataset displays a sharp increase (cf.\ Fig.~\ref{fig:proteins}a). This value is also close to that at which the size of the giant component of the network has a sharp increase and is consistently lower than the percolation threshold (cf.\ Fig.~\ref{fig:proteins}b).

The naive counting of small patterns of contacts in the database of proteins  displays some qualitative features predicted from the cluster expansion (Fig.~\ref{fig:entropy}, with counting of pattens in log scale). First, as discussed in Sect.~\ref{sect:graphs}, cliques are not likely. We can find in real proteins fewer small cliques than other kinds of graph and there are not cliques of degree higher than four. Small paths displaying large entropy
(Fig.~\ref{fig:proteins}c) are quite abundant in real proteins. Moreover, squares and 3--edge stars are abundant and display rather large entropy. On the contrary, pentagons are found in real proteins but display a rather low entropy from the cluster expansion. Also, paths with increasing length are more and more present in real proteins, although their entropy roughly decreases with their length in Fig.\ref{fig:entropy}.

However, it should be considered that the different graphs enumerated in the protein data set are not mutually exclusive. For example, paths are present in essentially all other graphs; in fact, they are the most abundant, increasing combinatorially with the number of edges. On the other hand, the estimation of the free energy in Eq.~(\ref{eq:f}) requires the knowledge of the number $\tn$ of graphs found as disjoint components, which is not the overall number $n_{\gamma}$ of graphs in the native contact map.

The calculation of $\tn$ is not straightforward because their definition requires to specify the order in which they have to be enumerated (cf.\ Eq.~(\ref{eq:order})). Following the prescription of Eq.~(\ref{eq:f}), we defined a set $\Gamma^*$ of graphs ordered according to the associated value of $\tg$ and enumerated the graphs in that order, removing the nodes after they have been counted once (from left to right in Fig.~\ref{fig:proteins}d). 

The simplest graphs $\twonodes$ and $\drawpath{3}$ are the most abundant ones and correspond to a large entropy contribution (upper panel in Fig.~\ref{fig:entropy}). The graphs $\Ngon{3}$ and $\drawpath{4}$ are predicted to display large entropy but they are suppressed by our node removal strategy (they are not displayed in Fig.~\ref{fig:proteins}d). The graphs $\stargraph{3}{0.25}$ and $\Ngon{4}$ are correctly predicted as the most abundant ones after those already described. For the 4 most abundant graphs, remarkably, the counts found in natural proteins are larger than for random decoys (red bars in Fig.~\ref{fig:entropy}), obtained from the collapse of a protein model in which all atoms interact with the same energy \cite{Negri2021}. Finally, our strategy to avoid double counting of nodes allows (roughly) to find a similar behaviour as a function of length for the paths abundances in real proteins (Fig.~\ref{fig:proteins}d) and the cluster expansion entropy (Fig.~\ref{fig:entropy}, upper panel), including the reduced entropy loss for 6--paths.

One can argue that the absence of triangles is due to the fact that they are not compatible with the formation of hydrogen bonds, that is one of the most important stabilizing interactions in protein. On the contrary, squares and paths can be stabilized by hydrogen bonds.

\begin{figure}
    \begin{align*}
    &\text{(a)} \quad \vcenter{\hbox{\includegraphics[width=0.41\linewidth]{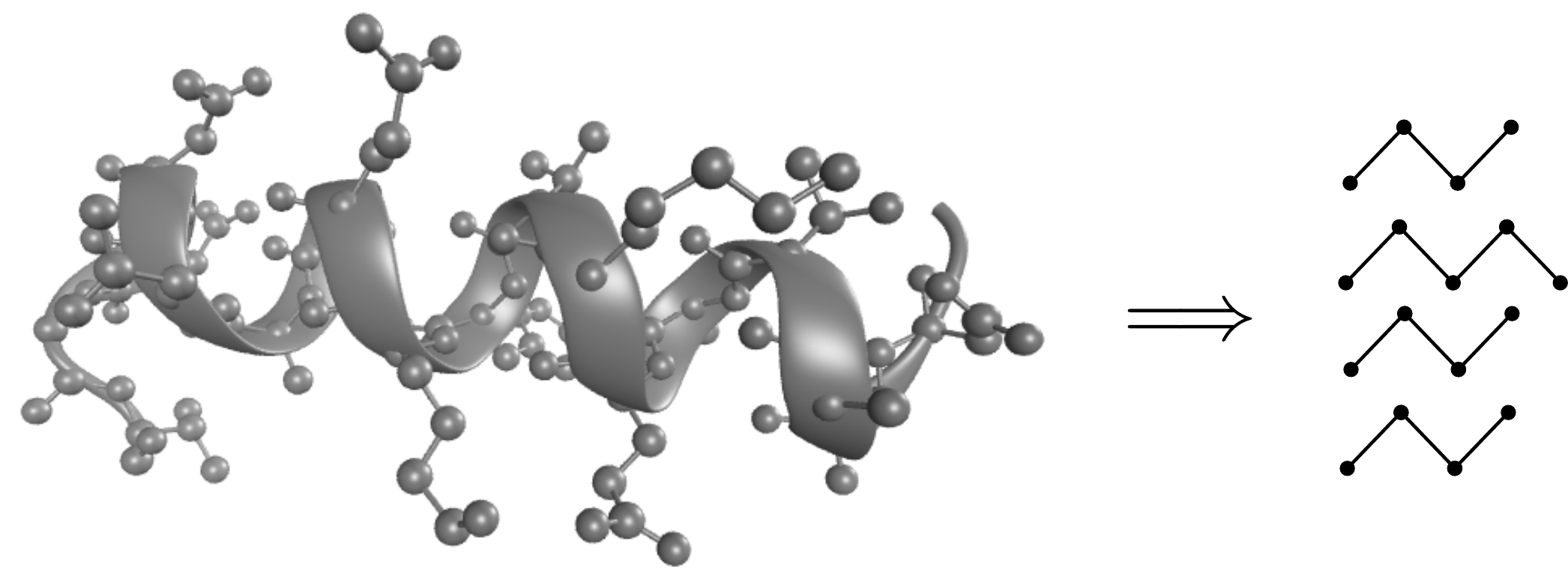}}}  \\
    &\text{(b)} \quad \vcenter{\hbox{\includegraphics[width=0.5\linewidth]{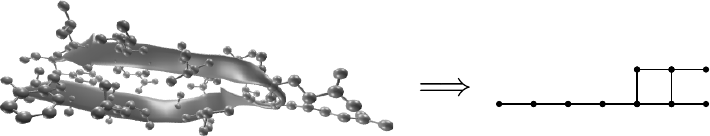}}} \\
    &\text{(c)} \quad \vcenter{\hbox{\includegraphics[width=0.52\linewidth]{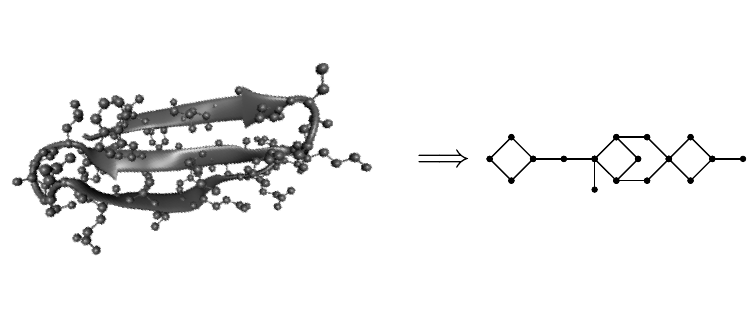}}} 
    \end{align*}
    \caption{The graphs arising from (a) the $\alpha$--helix of 1pgb protein, (b) the first $\beta$--hairpin of protein 1h4x and (c) the first 3-$\beta$--sheet of protein 3mmy.}
    \label{fig:secondary}
\end{figure}

It is also known that proteins are rich of secondary structures, most notably $\alpha$--helices and $\beta$--sheets. From the point of view of the graphs as defined above, 
both kind of structures are products of paths of various length, in the case of $\beta$--sheets sometimes decorated with squares (cf.\ Fig.~\ref{fig:secondary}).

Overall, paths and squares seem particularly abundant in proteins because associated with large entropy in sequence space.

\section{Conclusions}

A cluster expansion is a suitable tool to study the partition function of proteins.
In the case of the partition function in the space of sequences of proteins with a specified native state, the clusters have a very clear meaning of patterns of interactions defined by the geometry of the native conformation.

Proteins are dense, finite systems. For this reason, the cluster expansion has to be handled differently than in simpler diluted systems, like non--ideal gases. Here, one cannot expect that the terms of higher degree become negligible. This is not a problem for the convergence of the expansion, because it contains a finite set of terms, but makes the evaluation of the importance of the different clusters tricky. Finding the most important cluster corresponds to a huge and complicated graph spanning most of the amino acids of the protein would not be particularly useful. It would be as cumbersome as summing the full partition function and the results for a protein could be hardly generalized to other proteins. For this reason, we studied classes of clusters (e.g.,\ paths, polygons, etc.), comparing their relative importance. We also developed a sensible strategy to identify small important clusters in a connected network of interactions, solving the problem of avoiding a double--counting of the nodes.

Using a standard parametrization for the contact energies between amino acids, we found that clusters that contribute most to the partition functions are cycles with even numbers of nodes, while cliques are detrimental. These contributions can be partitioned into an energetic and an entropic part. Low--energy clusters are important for protein stability, high--entropy clusters for designability of the fold. As a rule, these two features are correlated. Small paths are highly entropic but poorly stabilizing. Stars and polygons are stabilizing but poorly entropic. Important exceptions are the square and the 6--path, that are both stabilizing and entropic.

We compared these results with the countingsw of different small contact patterns in real proteins. Overall, the kinds of patterns that are over-represented in real proteins tend to agree with those predicted to have large entropy. In particular, paths and squares are patterns particularly abundant in $\alpha$--helices and $\beta$--structures, respectively, that are the basic secondary structures of proteins.

Of course, one should consider that the thermodynamic stability is an epistatic constraint, but it is not the only factor that determines the evolutionary fitness of a protein. The formation of active sites, the kinetic accessibility, the need of avoiding aggregation are other requirements that could explain the differences between high--entropy graphs predicted by the model and those found in real proteins. Our findings can thus be regarded as the baseline on which evolution adds more stringent requirements.

\end{document}